\documentstyle[a4]{article}

\def\supp{\mathop{\rm supp}}
\def\sign{\mathop{\rm sign}}
\def\div{\mathop{\rm div}}
\def\Lie{{\pounds}}
\def\d{{\rm d}}
\def\Tor{{\tau}}
\def\Tr{\mathop{\rm Tr}}
\def\split#1{\vadjust{\kern#1}}

\begin{document}

\title{{\normalsize
\hfill{\rm ADP 96-41/M49} \\
\hfill{\rm gr-qc/9701046} \\[5mm]}
\bf Tensor distributions on signature-changing space-times
}

\author{
David Hartley\\
\it Dept.\ of Physics and Mathematical Physics\\
\it University of Adelaide, Adelaide, SA 5005, AUSTRALIA \\
{\tt DHartley{\rm @}physics.adelaide.edu.au} \\
\and
Robin W. Tucker\\
\it School of Physics and Chemistry\\
\it Lancaster University, Lancaster LA1 4YB, UK \\
{\tt R.W.Tucker{\rm @}lancaster.ac.uk} \\
\and
Philip A. Tuckey\\
\it Observatoire de Besan{\c c}on\\
\it Universit{\'e} de Franche-Comt{\'e}, 25010 Besan{\c c}on Cedex, FRANCE\\
{\tt pat{\rm @}obs-besancon.fr} \\
\and
Tevian Dray\\
\it Department of Mathematics\\
\it Oregon State University, Corvallis, OR  97331, USA \\
{\tt tevian{\rm @}math.orst.edu}
}

\date{15th January 1997}

\maketitle

%\widetext

%\pacs{04.20.Cv, 04.20.Me, 11.30.-j, 02.40.Hw}

\begin{abstract}
Irregularities in the metric tensor of a signature-changing space-time
suggest that field equations on such space-times might be regarded as
distributional. We review the formalism of tensor distributions on
differentiable manifolds, and examine to what extent rigorous meaning can
be given to field equations in the presence of signature-change, in
particular those involving covariant derivatives. We find that, for both
continuous and discontinuous signature-change, covariant differentiation
can be defined on a class of tensor distributions wide enough to be
physically interesting.
\end{abstract}

\section{Introduction}

A classical signature-changing space-time $M$ has a metric tensor $g$ whose
signature is Lorentzian in some regions and Euclidean in others.  Whenever
two regions of different signature exist on a connected component of $M$,
there must be some surface at which the metric tensor is either degenerate
(implying a singularity in the inverse metric) or discontinuous. In order
to give meaning to field equations involving the metric on such
space-times, either severe restrictions must be placed on the allowed class
of fields, such as assuming they be class $C^2$ \cite{Hayward} (see also
\cite{DrayHellaby}), or a distributional point of view must be adopted
\cite{DrayManogueTucker} (see also \cite{Hayward2,DrayManogueTucker2}.

In this note, we investigate to what extent the formalism of tensor
distributions can be applied to signature-changing space-times. Our main
conclusion is that rigorous meaning can be given to field equations for a
class of fields wide enough to be physically interesting.

Of course, tensor distributions on manifolds are by no means new: admirable
treatments of the topic can be found in many places ({\it eg\/}
\cite{Lichnerowicz,deRham,AMP}). Indeed, the formalism is quite standard in
studying shock waves in general relativity
\cite{Israel,ChoquetBruhat}. However, in such applications, it is usual to
assume the metric is continuous and non-degenerate, with irregularities
first appearing at the level of the connection. Since either of these
conditions on the metric can be violated under signature-change, the
standard formalism should be re-examined in this new context.

The difficulty is not {\it defining} tensor distributions in the
presence of signature-change, since the space-time metric plays no part
in establishing the topological spaces of test functions and test
tensors on a manifold. Rather, problems arise when tensor distributions
are associated with locally integrable tensors, since the volume element
of $g$ is usually used for this purpose. Under signature-change, this
may be degenerate, so an alternative volume element must be introduced.
Similarly, the volume element and connection are often used to define
differentiation of distributions, so care is needed here too. More
problematic is the definition of covariant differentiation, since both
the tensor fields and the connection components are distributional
in nature. 

We begin in sections \ref{Distributions} and \ref{Differentiation} with
a brief review of the standard treatment of distributions on manifolds,
highlighting the choices appropriate to signature-change. Distributions
related to hypersurfaces are discussed in section \ref{Hypersurface
distributions}.  Section \ref{Signature-change} then contains
applications to signature-changing space-times, including some simple
examples. We discuss our conclusions in section \ref{Discussion}. A more
detailed review of tensor distributions on degenerate space-times can be
found elsewhere \cite{Dray}.

\section{Distributions}\label{Distributions}

In this section, we review the standard formalism of tensor
distributions on differentiable manifolds, partly to establish our
notation. For missing details, we refer to standard works
\cite{Lichnerowicz,deRham,AMP}. Here, as elsewhere, we assume all
functions and tensors are smooth ({\it ie\/} $C^\infty$) unless
specifically mentioned. Similar definitions hold for lower degrees of
differentiability. 

Let $M$ be a paracompact $n$-dimensional differentiable manifold.  {\it
Test functions} on $M$ are (smooth) functions with compact support.  {\it
Scalar distributions} on $M$ are real or complex-valued continuous linear
functionals on the space of test functions, which is equipped with a
suitable topology. We denote the action of a distribution $D$ on a test
function $p$ by $D:p\mapsto D[p]$.

In a similar fashion, {\it test tensors} on $M$ are tensors with
compact support. The space of test tensors can be equipped with a suitable
topology by introducing an auxiliary {\it Riemannian} metric and
connection. However, the space-time metric is not required in this
definition. In fact, an equivalent topology can be defined without using
any metric at all \cite{ChoquetBruhat}. {\it Tensor distributions} are then
real or complex-valued continuous linear functionals on the topological
space of test tensors, and we write $T:U\mapsto T[U]$. As a linear
functional on test tensors, $T$ can be assigned a tensor type, which must
be dual to that of the test tensor $U$.

Multiplication of tensor distributions by functions, and tensor
products and contractions with ordinary tensors present no difficulty:
for a tensor distribution $T$, function $f$, vector $X$ and tensor
$S$,
\begin{eqnarray}
	(fT)[U] &=& T[fU]\nonumber\\
	(S\otimes T)[W\otimes U] &=& T[\langle S,W \rangle U]\nonumber\\
        T_X[U] &=& T[X\otimes U],
\end{eqnarray}
where $\langle S,W \rangle$ represents the total contraction of $S$ and $W$
(which has tensor type dual to that of $S$). Tensor products of tensor
distributions are not defined in general (just as for scalar products of
scalar distributions).

Given a local frame $\{X_a\}$ and dual coframe $\{\theta^a\}$ on
an open set $N$ in $M$, the {\it components} of $T$ are scalar
distributions defined in an obvious way.  For example, if $\alpha$ is a
covector distribution, then the component distributions $\alpha_a$ are
given by
\begin{equation}
	\alpha_a[p] = \alpha[pX_a]
\end{equation}
for any test function $p$ with $\supp p\subset N$. This leads to a local
expression for any tensor distribution $T$ in terms of its components and
frame and coframe elements, completely analogous to local
expressions of ordinary tensors. For example, on test vectors $V$ with $\supp
V\subset N$,
\begin{equation}
	\alpha[V] = \alpha_a\theta^a[V].
\end{equation}
The components transform in the usual way under changes of frame, and the
tensorial operations defined above have the usual expressions in terms of
components. In other words, tensor distributions can equally be regarded as
distribution-valued tensors.

Up to this point, no special structures (metric {\it etc}) have been
assumed on $M$. Now let $\omega$ be a volume element ({\it ie\/} a nowhere
vanishing $n$-form) on $M$. To every locally integrable tensor $S$ on $M$
we can associate a tensor distribution denoted $\widehat{S}$ by
\begin{equation}
	\widehat{S}[U] = \int_M{\langle S,U \rangle\omega}.
\end{equation}

It is possible to avoid the introduction of a volume element by using de
Rham currents \cite{deRham,Balasin}, and replacing test functions by test
$n$-forms. Tensor distributions can be defined in terms of their
components, and a distribution $\widehat{f}$ \split{1pt} associated with a
locally integrable function $f$ by $\widehat{f}[\phi] = \int_M{f\phi}$ for
any test $n$-form $\phi$. Here we shall follow the more conventional
approach.

\section{Differentiation}\label{Differentiation}

In order to define differentiation of distributions, we make further
use of the volume form. For an ordinary vector $X$ and a scalar
distribution $D$, we set
\begin{equation}\label{XD}
	(XD)[p] = -D[\div(pX)]
\end{equation}
for all test functions $p$, where the divergence is defined ({\it cf\/}
\cite{KobayashiNomizu}) by
\begin{equation}\label{div}
	\div(X)\omega = \Lie_X\omega = \d i_X\omega.
\end{equation}
The definition (\ref{XD}) is compatible with the action of $X$ on an ordinary
function: $X\widehat{f} = \widehat{Xf}$. In a coordinate chart $\{x^a\}$
with $\omega = k\d x^1\wedge ... \wedge\d x^n$ (where $k$ is a function in
general), we have the partial derivatives (in agreement with
\cite{ChoquetBruhat})
\begin{equation}
	\partial_a D[p] = -D[k^{-1}\partial_a(kp)].
\end{equation}
It is interesting to note that if a non-smooth volume element $\omega$ were
used then distributions would be only finitely differentiable, in contrast
with the usual presentation of distributions on ${\bf R}^n$. 

Other differential operators on tensor distributions are built from this
basic definition in the same fashion as for ordinary tensors. For example,
the exterior derivative $\d$ is uniquely prescribed by the requirements
that it be an anti-derivation of degree 1 on exterior form
distributions, that $\d^2=0$, and that $\d D$ for a scalar
distribution $D$ satisfy $i_X\d D = XD$ for ordinary vectors $X$. This
leads to $\d D[pX] = i_X\d D[p] = XD[p] = -D[\div(pX)]$ or, for any test
vector $V$,
\begin{equation}\label{dD}
	\d D[V] = -D[\div V].
\end{equation}

Likewise, covariant differentiation of tensor distributions is completely
determined by equation (\ref{XD}) and the usual properties of covariant
differentiation upon fixing the {\it connection map} $\nabla$ from vector
distributions to type (1,1) tensor distributions, satisfying the usual
requirements
\begin{eqnarray}
	\nabla(Y+Z) &=& \nabla Y + \nabla Z\\
        \nabla(D X) &=& \d D\otimes X + D\nabla \hat{X}\label{implicit}
\end{eqnarray}
for all vector distributions $Y$, $Z$, scalar distributions $D$ and
ordinary vectors $X$. Explicit formulae for covariant derivatives may be
written in terms of local components.  For example, if $Z$ is a vector
distribution with local coordinate components $Z^a$, then
\begin{equation}\label{cpt}
	(\nabla Z)_b{}^a \equiv \nabla_b Z^a 
        	= \partial_b Z^a + \Gamma^a{}_{bc} Z^c.
\end{equation}
Implicit in condition (\ref{implicit}) is the requirement that the product
$D \nabla \hat{X}$ be well-defined. Typically, the connection map on
distributions arises from a smooth connection, also denoted $\nabla$, on
$M$ ({\it ie\/} $\nabla X$ is smooth for all smooth vectors $X$). Setting
$\nabla\widehat{X} = \widehat{\nabla X}$ for all smooth vectors $X$
consistently determines the distributional connection map, since $D \nabla
\hat{X}$ is then well-defined for all scalar distributions $D$.

Given a smooth connection $\nabla$ on $M$, it is straightforward to show
that the divergence defined in equation (\ref{div}) and the alternative
definition $\nabla\cdot X = \Tr(\nabla X)$ satisfy
\begin{equation}
	\div X = \nabla\cdot X
\end{equation}
for all vectors $X$ if, and only if 
\begin{equation}\label{metrictorsion}
	\nabla\omega + \Tr(\Tor) \otimes \omega = 0,
\end{equation}
where $\Tor$ is the type (1,2) torsion tensor of $\nabla$. If
these conditions are met, then
\begin{equation}\label{XD1}
	(XD)[p] = -D[\nabla\cdot(pX)]
\end{equation}
can be used as an equivalent starting point for the differentiation of
distributions, leading to an elegant expression for the absolute
derivative:
\begin{equation}\label{DT1}
	\nabla T[U] = -T[\nabla\cdot U],
\end{equation}
in which the first argument of $U$ must be a covector, and $\nabla\cdot U$
is the trace over the first two arguments of $\nabla U$. However, we are
interested in the possibility that $\nabla$ not be smooth (see below), let
alone satisfy condition (\ref{metrictorsion}), so we forego (\ref{XD1}) and
(\ref{DT1}), and adhere to our choice of definition (\ref{XD}).

If $\nabla$ is not smooth, by which we mean that it is not associated with
a smooth connection on $M$, then the covariant derivative may not be
defined for all tensor distributions. In particular, products such as
$D\nabla X$ from property (\ref{implicit}) will not be defined for all
$D$ and $X$. The same holds for terms like $\Gamma^a{}_{bc} Z^c$
{}from the component version (\ref{cpt}). However, by restricting the class
of tensor distribution $T$ being differentiated with respect to a given
$\nabla$, it may still be possible to give meaning to $\nabla T$.

\section{Hypersurface distributions}\label{Hypersurface distributions}

Since we are interested in discontinuities and singularities at a change of
signature, it is worth examining distributions with support on and between
regions of different signature. For our purposes, it suffices to suppose
$M$ is orientable and divided into two disjoint open regions $M^+$ and
$M^-$ by an $(n-1)$-dimensional submanifold $\Sigma$, which in turn is
defined by the equation $\lambda=0$ for some function $\lambda$ on $M$
satisfying $\d\lambda\not=0$ in a neighbourhood of $\Sigma$. We fix an
orientation by taking $\Sigma = \partial M^-$.

The
Heaviside scalar distributions $\Theta^\pm$ are defined by
\begin{equation}
	\Theta^\pm[p] = \int_{M^\pm}{p\omega}
\end{equation}
for any test function $p$, while the Dirac 1-form distribution $\delta$ is
\begin{equation}
	\delta[V] = \int_\Sigma{i_V\omega}
\end{equation}
where $V$ is any test vector.

Introducing a {\it Leray form} $\sigma$ in a neighbourhood of $\Sigma$,
defined by 
\begin{equation}\label{Leray}
	\omega = \d\lambda\wedge\sigma
\end{equation}	
allows us to define the usual scalar Dirac distribution
$\delta(\lambda)$ as
\begin{equation}
	\delta(\lambda)[p] = \int_\Sigma{p\sigma}.
\end{equation}
The right-hand side is independent of the choice of $\sigma$ satisfying
(\ref{Leray}), but depends on the choice of the function $\lambda$ used
to describe $\Sigma$. The scalar $\delta(\lambda)$ and
($\lambda$-independent) 1-form $\delta$ are related by
\begin{eqnarray}
	\delta[V] 
	   &=& \int_\Sigma{\left((i_V\d\lambda)\sigma 
			   - \d\lambda\wedge i_V\sigma\right)}\nonumber\\
	   &=& \delta(\lambda)[\langle \d\lambda,V \rangle]
\end{eqnarray}
since $\d\lambda$ vanishes under pullback to $\Sigma$. Thus
\begin{equation}
	\delta = \delta(\lambda)\d\lambda.
\end{equation}
The usual scaling law $\delta(a\lambda) = \frac{1}{a}\delta(\lambda)$
is simply an expression of the fact that $\delta(\lambda)$ transforms
as the component of a 1-form.

Applying the exterior derivative (\ref{dD}) to the Heaviside
distribution $\Theta^+$ we have, for any test vector $V$,
\begin{equation}
	\d\Theta^+[V] = -\Theta^+[\div V]
		= -\int_{M^+}{\d i_V\omega}
		= -\int_{\partial M^+}{i_V\omega}
		= \delta[V]
\end{equation}
since $\partial M^+ = -\Sigma$. A similar calculation can be done for
$\Theta^-$, so we finally have the satisfying result
\begin{equation}\label{dTheta}
	\d\Theta^\pm = \pm\delta.
\end{equation}

A function $f$ on $M$ is {\it regularly $C^k$ discontinuous} at $\Sigma$ if
$f$ and its first $k$ derivatives are continuous on $M^\pm$ and converge
uniformly to limits $f^\pm_\Sigma$ {\it etc\/} at $\Sigma$. We will
consider only regularly $C^\infty$ discontinuous functions and suppress the
degree of differentiability. A regularly discontinuous tensor $S$ is one
whose components in any given chart intersecting $\Sigma$ are regularly
discontinuous functions. The {\it discontinuity} $[\![S]\!]$ of $S$ is an
ordinary continuous tensor over $\Sigma\subset M$ defined by
\begin{equation}
	[\![S]\!] = S^+_\Sigma - S^-_\Sigma.
\end{equation}

Let $\widehat{f}$ be the distribution associated with a regularly
discontinuous function $f$. Then $\widehat{f}$ can be written
\begin{equation}\label{hatf}
	\widehat{f} = \Theta^+f^+ + \Theta^-f^-
\end{equation}
where $f^\pm$ are arbitrary smooth extensions of $f|_{M^\pm}$ to $M$. In
particular, $f^\pm|_\Sigma = f^\pm_\Sigma$. It follows
that
\begin{equation}
	\d\widehat{f} = \Theta^+\d f^+ + \Theta^-\d f^- + [\![f]\!]\,\delta,
\end{equation}
where the last term is well-defined because $[\![f]\!]$ is continuous on
$\supp\delta = \Sigma$.  Expression (\ref{hatf}) extends in the obvious way
to regularly discontinuous tensors, and we have, for smooth $\nabla$ and
regularly discontinuous $S$
\begin{equation}\label{DsmoothS}
	\nabla\widehat{S} = \Theta^+\nabla S^+ + \Theta^-\nabla S^- 
		+ \delta\otimes[\![S]\!].
\end{equation}

At this point, we wish to consider two possibilities for defining a
connection on hypersurface distributions which is itself no longer smooth. As
mentioned at the end of section \ref{Differentiation}, such a connection
may not be defined on all tensor distributions.

First, suppose $\nabla$ is regularly discontinuous, by which we mean that
$\nabla X$ is regularly discontinuous for smooth vectors $X$. Then it is
not possible to make sense of condition (\ref{implicit}) when $D$ contains
a $\delta(\lambda)$ factor. However, if $Y$ is a locally integrable vector,
then it can be written as a sum of terms of the form $fX$ where $f$ is
locally integrable and $X$ is smooth. So $\nabla\widehat{Y}$ contains terms
$\nabla(\widehat{f}\,X) = \d\widehat{f}\,\otimes X + \widehat{f}\,\nabla X$
in which $\nabla X$ is regularly discontinuous and hence
$\widehat{f}\,\nabla X$ is well-defined. The extension to locally
integrable tensors $S$ is straightforward. If $S$ is regularly
discontinuous, then equation (\ref{DsmoothS}) holds in the slightly
modified form
\begin{equation}\label{DregdiscS}
	\nabla\widehat{S} = \Theta^+\nabla^+S^+ + \Theta^-\nabla^-S^- 
		+ \delta\otimes[\![S]\!],
\end{equation}
where $\nabla^\pm$ are arbitrary smooth extensions of $\nabla|_{M^\pm}$
to $M$.

Secondly, we consider the possibility of defining a distributional
connection which is not only discontinuous, but contains a Dirac $\delta$
part. In this case, it will not be possible to make sense of condition
(\ref{implicit}) even when $D$ is merely discontinuous. Accordingly, we
restrict our attention to those distributions associated with smooth
tensors.  As described earlier, fixing the connection map on vector
distributions consistently determines the covariant derivative of other
tensor distributions. For a vector $X$ we postulate a connection map of the
form
\begin{equation}\label{nabladelta}
	\nabla\widehat{X} = \overline{\nabla} \widehat{X} 
				+ \delta\otimes k(X),
\end{equation}
where $\overline\nabla$ is some regularly discontinuous connection and $k$
is a fixed $T(M)$-valued function, whose properties are to be
determined. The Leibniz rule (\ref{implicit}) applied to both $\nabla$ and
$\overline\nabla$ implies that, for any function $f$, 
\begin{equation}\label{klinear}
	\delta\otimes k(f X) = f \delta\otimes k(X),
\end{equation}
which is satisfied if $k$ is a type (1,1) tensor. With this condition on
$k$, the connection map given by (\ref{nabladelta}) defines a connection
$\nabla$ on tensor distributions associated with smooth tensors. Extension
of definition (\ref{nabladelta}) to $\nabla\widehat{Y}$ with $Y$ regularly
discontinuous is ruled out by the same argument: $Y$ is a sum of terms $fX$
where $f$ is regularly discontinuous and $X$ is smooth, so condition
(\ref{klinear}) still applies. However, the right-hand side is undefined
for general $f$ ({\it eg\/} $f=\Theta^\pm$) unless $\delta\otimes k(X) = 0$
for $X$ smooth, which reduces definition (\ref{nabladelta}) to
$\nabla\widehat{Y} = \overline{\nabla} \widehat{Y}$.

\section{Signature-change}\label{Signature-change}

We now consider $M$ to be a signature-changing space-time, with $\Sigma$
the surface of signature-change in the metric $g$. Two types of
signature-change are considered here: {\it discontinuous}, in which $g$ is
regularly discontinuous at $\Sigma$ with $\det g^\pm_\Sigma \not= 0$, and
{\it continuous}, in which $g$ is smooth, but $\det g$ vanishes on
$\Sigma$.  We further demand that the induced metrics on $\Sigma$ from each
side agree and are non-degenerate. With these assumptions, $g$ can be
written
\begin{equation}
	g = N\d\lambda\otimes\d\lambda + h,
\end{equation}
where $N$ changes sign at $\Sigma$, being regularly discontinuous with
$N_\Sigma^\pm \not= 0$ in one case, and smooth with $N|_\Sigma=0$ in the
other.

The Levi-Civita connection for the restrictions of this metric to the open
sets $M^\pm$ can be computed as usual. In particular, in adapted
coordinates such that $\partial_\lambda$ is normal to the surfaces of constant
$\lambda$ near $\Sigma$, we have
\begin{equation}\label{Gamma000}
	\left.
	\d\lambda(\nabla_{\partial_\lambda}{\partial_\lambda})
	\right|_{M^\pm} = \frac{\partial_\lambda N}{2N}.
\end{equation}

For discontinuous signature-change, this allows us to construct a
regularly discontinuous connection $\nabla$ based on arbitrary smooth
extensions $\nabla^\pm$ to $M$ of the Levi-Civita connections on
$M^\pm$. From section \ref{Hypersurface distributions}, we know that
such a connection can be applied to those distributions associated with
locally integrable tensors, including regularly discontinuous
tensors. Hence, it is possible to discuss first order field equations for this
class of fields, or second order field equations for smooth tensor
fields. The metric tensor is included in the class of covariantly
differentiable tensor distributions, so the metric compatibility of
$\nabla$ can be established from equation (\ref{DregdiscS}) as
\begin{eqnarray}\label{nablag}
	\nabla \widehat{g} &=& \Theta^+\nabla^+g^+
		 + \Theta^-\nabla^-g^-
		 + \delta\otimes [\![g]\!]\nonumber\\
		 &=& \delta\otimes [\![g]\!].
\end{eqnarray}
Thus the (metric-compatible) Levi-Civita connections on $M^\pm$ give rise
to a non-metric compatible connection $\nabla$ for distributions on $M$! To
rectify this, we might be tempted to add a Dirac $\delta$ term to $\nabla$,
as discussed in section \ref{Hypersurface distributions}, but this would
make $\nabla \widehat{g}$ undefined, since $g$ is not smooth.

As a simple concrete example of discontinuous signature-change, consider
the metric tensor
\begin{equation}\label{discg}
	g = \sign(t)\d t\otimes\d t + a(t)^2\d x\otimes\d x
\end{equation}
on $M={\bf R}^2$, where $a > 0$. In this particular case, the metric volume
forms on $M^\pm$ extend to give a smooth volume form $\omega = a\d
t\wedge\d x$, so this is a natural choice. The non-zero Levi-Civita
connection components $\Gamma^a{}_{bc} = \d x^a(\nabla_{\partial_b}
\partial_c)$ ($a,b,c=t,x$) for $g$ restricted to $M^\pm$ are
\begin{equation}
	\Gamma^t{}_{xx}|_{M^\pm} = \mp a\dot{a}\qquad
	\Gamma^x{}_{xt}|_{M^\pm} = \frac{\dot{a}}{a},
\end{equation}
and these expressions extend smoothly to the whole of $M$, determining
a regularly discontinuous connection $\nabla$. As anticipated in
equation (\ref{nablag}), the connection is not metric compatible, since
(using $\Gamma^{\pm\,t}{}_{tt} = 0$)
\begin{equation}
	\nabla_t g_{tt} = \partial_t g_{tt} = 2\delta(t).
\end{equation}

For continuous signature-change, we can take $N = \lambda$ near $\Sigma$
if we further assume $\d N\not=0$ on $\Sigma$ \cite{KossowskiKriele}. Then
the right-hand side of equation (\ref{Gamma000}) simplifies to give
\begin{equation}
	\left.
	\d\lambda(\nabla_{\partial_\lambda}{\partial_\lambda})
	\right|_{M^\pm} = \frac{1}{2\lambda},
\end{equation}
which is singular on $\Sigma$. Hence we cannot construct a regularly
discontinuous connection from the Levi-Civita connections on $M^\pm$.
Nonetheless, the connection can be given a meaning directly, and for {\it
all} distributions. This is most easily seen from the component expression
(\ref{cpt}) and the fact that $\frac{1}{\lambda}D$ can be defined for any
distribution $D$ (see {\it eg\/} \cite{AMP}).  However, this can be done in
many ways, as the distribution $E$ satisfying $\lambda E = D$ is determined
only up to addition of a term proportional to $\delta(\lambda)$. Up to such
ambiguities, it is possible to discuss field equations of any order for any
tensor distributions within continuous signature-change. The question of
metric compatibility can again be posed, and here, the natural connection
may be metric compatible.

A simple example of continuous signature-change, close to (\ref{discg}) above
for the discontinuous case, is given by the metric tensor
\begin{equation}
	g = t\d t\otimes\d t + a(t)^2\d x\otimes\d x
\end{equation}
on $M={\bf R}^2$, where again $a > 0$. The metric volume forms on $M^\pm$
vanish on $\Sigma$, so an alternative must be introduced: $\omega = 
a\d t\wedge\d x$ is an obvious choice in these coordinates. The non-zero
Levi-Civita connection components for $g$ restricted to $M^\pm$ are
\begin{equation}
	\Gamma^t{}_{tt}|_{M^\pm} = \frac{1}{2t}\qquad
	\Gamma^t{}_{xx}|_{M^\pm} = -\frac{a\dot{a}}{t}\qquad
	\Gamma^x{}_{xt}|_{M^\pm} = \frac{\dot{a}}{a},
\end{equation}
giving rise to singularities at $t=0$. As discussed above, the connection
map determined by these expressions is defined for general vector
distributions, fixed up to addition of terms proportional to
$\delta(t)$. In $\nabla\,\widehat{g}$, these terms are annulled by the smooth
function $g_{tt}=t$, since $t\delta(t) = 0$. A simple calculation verifies
that the connection is metric compatible.

\section{Discussion}\label{Discussion}

We began with a review of the standard theory of tensor distributions on
manifolds, emphasising that the space-time metric need not be used in
establishing the formalism. We chose definitions for ordinary and covariant
differentiation which depend on an auxiliary volume element rather than on a
full metric tensor, and which separate the role of this auxiliary volume
element from that of the connection components. This avoids introducing
unnecessary irregularities when the metric tensor and connection are not
smooth.

We then considered distributions associated with a hypersurface, detailing
how the Dirac $\delta$ distribution may conveniently be regarded as an
exact 1-form, $\delta = \d\Theta$, where $\Theta$ is a Heaviside
distribution, and discussing the important class of distributions arising
{}from regularly discontinuous tensors. Extensions of the covariant
derivative to cases where the connection components are not smooth were
described, and shown to be well-defined, at least on a restricted class of
distributions.

Finally, we applied the formalism to signature-changing metrics, both
continuous and discontinuous, and showed to what extent we can make
rigorous sense of covariant differentiation, and thus of tensor field
equations. A surprising result for discontinuous signature-change is that
the Levi-Civita connections on the Lorentzian and Euclidean regions extend
to include the hypersurface in such a way that the overall connection is no
longer metric-compatible! Einstein's equations in this setting have been
discussed from a variational approach in \cite{Dray2}.

The main obstruction to extending covariant differentiation with respect to
a connection based on the metric to {\it all\/} distributions is the
occurrence of undefined products of distributions. This deficiency might be
remedied by resorting to the Colombeau algebra of distributions, as has
been recently proposed for cosmic strings \cite{Clarke} and for other
distributional sources \cite{Balasin}.

\section*{Acknowledgments}

It is a pleasure to thank Charles Hellaby, Corinne Manogue, and J{\"o}rg Schray
for helpful discussions.  TD would like to thank the School of Physics \&
Chemistry at Lancaster University and the Department of Physics and
Mathematical Physics at the University of Adelaide for kind hospitality
during his sabbatical visits.  This work was partially supported by NSF
Grant PHY-9208494 (TD), a Fulbright Grant under the auspices of the
Australian-American Educational Foundation (TD), the Graduate College on
Scientific Computing, University of Cologne and GMD St Augustin, funded by
the Deutsche Forschungsgemeinschaft (DH), the Human Capital and Mobility
Programme of the European Union (RWT), and the Alexander von Humboldt
Foundation (PAT).

\end{document}